\documentclass[preprint,showpacs,preprintnumbers,amsmath,amssymb]{revtex4}
\usepackage{graphicx}

\begin{document}

\thispagestyle{empty}

\title{Demonstration of the asymmetric lateral Casimir
force between corrugated surfaces in the nonadditive regime}

\author{H.-C.~Chiu,${}^1$
G.~L.~Klimchitskaya,${}^2$
V.~N.~Marachevsky,${}^3$
V.~M.~Mostepanenko,${}^4$
and U.~Mohideen${}^1$}

\affiliation{${}^1$Department of Physics and Astronomy,
University of California,
Riverside, California 92521, USA \\
${}^2${North-West Technical University,
Millionnaya St. 5, St.Petersburg,
191065, Russia}\\
${}^3$V.A.~Fock Institute of Physics, St.Petersburg State
University, St.Petersburg, 198504, Russia \\
${}^4${Noncommercial Partnership ``Scientific Instruments'',
Tverskaya St. 11, Moscow,
103905, Russia}}

\begin{abstract}
The measurement of the lateral Casimir force between two aligned
sinusoidally corrugated Au-coated surfaces has been performed in
the nonadditive regime. The use of deeper corrugations also allowed 
to demonstrate an asymmetry in the phase dependences of the lateral 
Casimir force, as  predicted earlier. The measurement data are
found to be in excellent agreement with the exact theoretical 
results computed at T=300 K including effect of real material 
properties. The deviations between the exact theory and the 
proximity force approximation are quantified. The obtained results 
are topical for applications in nanomachines.
\end{abstract}
\maketitle

The most universally known {\it normal} Casimir force \cite{1}
leads to an attraction between two closely spaced bodies
directed perpendicular to their surfaces.
This attraction arises due to zero-point and thermal fluctuations
of the electromagnetic field. The phenomenon is  important
for numerous fields ranging from extra-dimensional physics
to nanotechnology and has recently attracted considerable
experimental and theoretical attention (see Ref.~\cite{2}).
The most intriguing feature of the Casimir force is its
nonadditivity and thus the complicated dependence on the shape of the
boundary surfaces connected with diffraction effects.
The nontrivial behavior of the normal
Casimir force was experimentally demonstrated in the configuration
of a smooth sphere above a sinusoidally corrugated plate \cite{3}
in the additive regime. Recently the normal Casimir force between an Au
coated sphere and a Si plate with an array of rectangular corrugations
was measured in the nonadditive regime \cite{4} (where the corrugation
period $\Lambda$ is smaller than the mean separation $z$ between
the surfaces \cite{5}). The measurement data were shown to deviate
from the additive theoretical results. These deviations, however, are
smaller by about 50\% than the prediction of Ref.~\cite{6} for ideal
metals which might be explained by the interplay between nonzero skin
depth and geometry effects \cite{4}. Exact calculations
performed at $T=0$ taking into account the nonzero skin depth
\cite{7} favor this conjecture.
In the configuration of two sinusoidally corrugated bodies (a sphere and a
plate) there also arises the {\it lateral}
Casimir force \cite{8,9} which gives the possibility to actuate
lateral translations in micromachines by means of the electromagnetic
zero-point fluctuations. In \cite{9} this force was measured in
the additive regime ($\Lambda\gg z$). It was shown to oscillate
sinusoidally as a function of the phase shift between the two
corrugations.

In this Rapid Communication we report the first measurement of the 
lateral Casimir
force between two aligned sinusoidally corrugated surfaces of a
sphere and a plate in the nonadditive regime where $\Lambda\sim z$.
The measurement data are found to be in excellent agreement with
the exact computational results at $T=300\,$K taking into
account real material properties. The deviation of both the experimental
data and the exact theory from the prediction of the proximity force
approximation (PFA) is quantified. The use here of much deeper corrugations
than in Ref.~\cite{9} enabled the first demonstration of the asymmetry of
the lateral force predicted in \cite{9}.
Below we present the obtained experimental results and their comparison
with theory by describing only the additional features of this experiment,
as compared with Ref.~\cite{9}.

The experimental setup
is shown in Fig.~1. A chamber with
a pressure less than 10\,mTorr 
 contains the sinusoidally corrugated Au-coated grating
of size $5\times 5\,\mbox{mm}^2$ vertically mounted on the piezotube of
an atomic force microscope (AFM). The corrugations have an average
period $\Lambda=574.7\,$nm (more than two times smaller than in \cite{9}
in order to achieve the nonadditive regime)
and an amplitude of $A_1=85.4\pm 0.3\,$nm (59\,nm in \cite{9}).
This grating was used as the
first test body. A $320\,\mu$m long V-shaped silicon nitride cantilever
of the AFM was uniformly coated with 40\,nm of Al to improve its
thermal and electric conductivity and prevent deformation due to differential
thermal expansion in vacuum. Then  a $200\pm 4\,\mu$m diameter
polystyrene sphere was placed at the end of the cantilever with
conductive Ag epoxy. Next, a freshly cleaved mica sheet of
$400\,\mu$m length, $200\,\mu$m width and a few micrometer thickness
was attached to the bottom of the sphere also with Ag epoxy.
One more similar sphere was then attached to the bottom, free end of
the mica sheet (see Fig.~1). This last sphere was used as the second
test body. The silver epoxy is rigid at all attachments. The resulting
system was uniformly coated \cite{9a}
with a 10\,nm layer of Cr and then with a 50\,nm
layer of Au in a thermal evaporator.

The lateral Casimir force arises between two perfectly aligned uniaxially
corrugated test bodies of the same $\Lambda$ \cite{5,8,9}.
To imprint corrugations on the sphere, the grating
was used as a template. It had a 300\,nm Au coating applied on top of
sinusoidal corrugations made of hard epoxy on a 3\,mm thick pyrex
substrate. In contrast to Ref.~\cite{9} where a soft plastic grating
coated with Al was used, here, the hard epoxy does not require the
deposition of the Al layer. The smaller $\Lambda$
required the use of a precise stepper motor and piezo-controlled
imprinting technique. At first, the sphere was brought
in contact with the grating using a micromanipulator.
Then a hard flat surface was moved in $10\,\mu$m steps using
a stepper motor till it comes into contact with the other side of
the sphere which is sandwiched against the grating. Next the imprinting
was done by applying a voltage to the $z$-piezo to squeeze the sphere
between the grating and the hard flat surface. To obtain
deeper imprints preserving sphericity
additional pressure was put on the sphere using the
stepper motor. With the help of the same motor the hard flat surface
was gently removed and the sphere was then moved horizontally to a
different position on the grating (as the latter might have
changed its local amplitude during the imprint process).
The amplitude of the imprinted corrugations over a
$30\times 30\,\mu\mbox{m}^2$ area
on the sphere was
measured with an AFM to be $A_2=13.7\pm 0.4\,$nm (60\% increase as
compared with \cite{9}). The diameter of the Au-coated sphere was
measured to be $2R=194\pm 0.3\,\mu$m using a scanning electron
microscope. The corrugations were
examined using an AFM and found to be homogeneous.

The distinctive feature of the setup shown in Fig.~1 is that
a lateral force along the $x$-axis tangential to the corrugated sphere and
a grating would lead to the vertical bending of the cantilever
(measured using the bi-cell photodiodes A and B in Fig.~1), whereas a
force acting normal to the test bodies would lead to a torsional
deflection. The torsional spring constant was found to be 46 times
larger than the bending spring constant using an independent quad
cell AFM measurement of the torsional movement of such a cantilever.
Thus, the normal Casimir force could lead to only a negligible
change in the second sphere position and the phase of
the corrugations.
First, the torsional cantilever deflection due to voltages $V$
applied to the grating while the sphere remained grounded
was measured by means of the difference signal
between the bi-cell photodiodes. The measured deflection
signal $S$ at $z=1\,\mu$m, where the Casimir force is negligible, was
fitted to the following expression for the normal electrostatic force
between the sphere and the grating:
\begin{eqnarray}
&&
F_{\rm nor}^{\rm el}(z,\varphi)\!=k_{\rm tor}S_{\rm nor}^{\rm el}\!\!=
-2\pi\epsilon_0(V-V_0)^2\!\left[
\frac{R}{2z}\,\frac{1}{\sqrt{1-\beta^2}}\right.
\nonumber \\
&&~
+c_0+c_1\frac{z}{R}+
c_2\frac{z^2(2+\beta^2)}{2R^2}+
c_3\frac{z^3(2+3\beta^2)}{2R^3}
\nonumber \\
&&~
+c_4\frac{z^4(8+24\beta^2+3\beta^4)}{8R^4}+
c_5\frac{z^5(8+40\beta^2+15\beta^4)}{8R^5}
\nonumber \\
&&~+\left.
c_6\frac{z^6(16+120\beta^2+90\beta^4+5\beta^6)}{16R^6}
\vphantom{\frac{1}{\sqrt{1-\beta^2}}}
\right].
\label{eq1}
\end{eqnarray}
\noindent
Here, $k_{\rm tor}$ is the normal force calibration constant,
$V_0$ is the residual potential difference between the grating
and the
sphere when both are grounded,
$c_i$ are the numbers listed in Ref.~\cite{10},
$\beta=(A_1^2+A_2^2-2A_1A_2\cos\varphi)^{1/2}/z$, and
$\varphi=2\pi x/\Lambda$ is the
phase shift between the corrugations on both surfaces.
Equation (\ref{eq1}) was derived using the exact formula for the electric
force in the sphere-plate geometry \cite{10} and taking the
corrugations into account by means of the PFA. We have checked that within the
measurement range, the analytical result (\ref{eq1}) coincides
up to 2.8\% with the
 numerical solution of the Poisson
equation using the software Comsol Multiphysics
(http://www.comsol.com).
The good agreement is explained by the absence of
diffraction effects for the static electric field. The measurement of
 $S$  was repeated for
eight different voltages between $-0.52\,$ to 0.47\,V. The mean value
of $V_0$ found from the fit to Eq.~(\ref{eq1}) is
$V_0=-39.6\pm 1.6\,$mV.

Then the cantilever bending due to the lateral Casimir force was
measured.
 In this case only the residual voltage $V_0$
was applied to the grating in order to make the electric force
equal to zero. The $x$-piezo was used to move the grating
along the $x$-axis and thus change $\varphi$.
The $z$-piezo, which was independently controlled by an external voltage
source, was used to change $z$.
The piezo extensions with applied voltage in both directions were
calibrated using optical interferometry \cite{11}. Small deviations
of the grating from the $x$-axis during the movement were corrected as
described in \cite{9}. Initially the corrugated sphere was positioned
3.79\,nm from the separation on
contact between the two surfaces $z_0$ determined by the corrugations and
the highest roughness peaks.
The mechanical
drift of this position was verified to be very small (average value of
0.14\,nm/min). A phase shift was introduced by moving the $x$-piezo
continuously for a total distance of $3.3\,\mu$m at 0.103\,Hz.
The photodiode signal corresponding to the cantilever deflection was
filtered with a low pass filter with a 30\,ms time constant and
recorded at each of the 8192 points corresponding to $x$-changes of
0.4\,nm.  Then the separation from $z_0$ was
increased by 3.6\,nm to 7.39\,nm and the deflection signal
$S_{\rm lat}^{\rm C}$  was similarly measured as a function of
$\varphi$ and recorded. After this, the separation from
$z_0$ was increased by 3.96\,nm to 11.35\,nm and the measurements
repeated. Next $S_{\rm lat}^{\rm C}$ due to
the lateral Casimir force as a function of
$\varphi$ was measured at separations of 20.05, 32.48,
45.30, 58.01 and 70.86\,nm from  $z_0$.

To convert the measurement data from $S_{\rm lat}^{\rm C}$  into
values of the lateral Casimir force at some specific
separation, it is necessary to determine the values of $k_{\rm ben}$
and $z_0$. This was achieved by measuring the cantilever deflection
signal due to the lateral electrostatic force which arises when a
voltage is applied to the grating. The measurements of this signal
was performed at small separations from close to $z_0$ to $z_0+120\,$nm.
First, a voltage of 141.456\,mV was applied to the grating.
The sphere was kept at a distance
of 3.96\,nm from $z_0$. The phase shift was again changed
continuously at a frequency
of 0.103\,Hz with the $x$-piezo to a maximum translation of $3.3\,\mu$m.
The cantilever deflection corresponding to the total
force (sum of the lateral electrostatic and lateral Casimir forces)
was recorded as 8192 evenly spaced data points.
The sphere was moved further away from the grating
by 5.40\,nm to a separation of 9.36\,nm and the measurement was repeated.
The cantilever deflection signal due to the total force was measured
at many other $z$ at the same voltage and also for
a second voltage of 101.202\,mV applied to the grating.
To obtain the deflection signal
due to the lateral electric force alone,
the previously measured  signal $S_{\rm lat}^{\rm C}$
was fitted to the sum of harmonics of the form
$A_k(z)\sin(k\varphi)$ with $1\leq k\leq5$
which takes into account the fact that the lateral
Casimir force is asymmetric.
After determination of the coefficients $A_k$, the signal
$S_{\rm lat}^{\rm C}$
was subtracted from the data for the total deflection
signal. As the sphere-grating separations in the measurements
of the lateral Casimir
force and of the total force are not identical, interpolation was
used to determine the values of $A_k$ at the separations corresponding to the
total lateral force. The obtained data for the lateral electrostatic force
were fitted to its analytic expression
$F_{\rm lat}^{\rm el}(z,\varphi)=k_{\rm ben}S_{\rm lat}^{\rm el}$
which is derived from Eq.~(\ref{eq1}) by the intregration with
respect to $z$ and subsequent differentiation with respect to $\varphi$.
As a result, the quantities $k_{\rm ben}$ and $z_0$ were found.
This was repeated for four different electrostatic force measurements and
the average values obtained are
$k_{\rm ben}=1.27\pm 0.06\,$nN per unit $S$
and $z_0=117.3\pm 3.0\,$nm (see Ref.~\cite{12a} for details of
electrostatic calibrations).

An independent control measurement of $V_0$ was done at $z=127.3\,$nm
using the parabolic dependence of the $F_{\rm lat}^{\rm el}$ on $V$.
Five different voltages close to the residual potential difference
were applied to the grating and the cantilever deflection was measured
as a function of $\varphi$ leading to
$V_0=-39.4\,$mV. This is consistent with $V_0=-39.6\,$mV found
earlier at large separations and confirms that the residual potential
difference is separation-independent.

The resulting lateral Casimir force
$F_{\rm lat}^{\rm C}=k_{\rm ben}S_{\rm lat}^{\rm C}$
at a separation $z=124.7\,$nm over four corrugation periods is shown
in Fig.~2 as dots versus the phase shift
[$x/\Lambda=\varphi/(2\pi)$] between the corrugated surfaces.
Similar results were obtained at
all $z$.
The mean values and the variances of the max$\,F_{\rm lat}^{\rm C}$
at each $z$ were calculated using the measurement
data for different periods.
The maximum values of the $F_{\rm lat}^{\rm C}$,
as a function of $z$, are shown
as crosses in Fig.~3. The arms of the crosses indicate the
errors determined at a 95\% confidence level using the procedure of
Ref.~\cite{10}. The error in the measurement of the absolute
separations,
 $\Delta z$, is a combination of the error in the
determination of $z_0$ indicated above and half of the first step of
the $z$-piezo $\Delta_{p}z\approx 2\,$nm. By combining these errors,
we obtain $\Delta z=4\,$nm (a factor of eight improvement as
compared with \cite{9}). The errors in the magnitudes of the
$F_{\rm lat}^{\rm C}$ were found by combining the random errors due to
the averaging over the different
periods and the systematic errors due to the uncertainty
in $k_{\rm ben}$ (all other systematic errors are negligible).
As an example, the total relative error of the
max$\,F_{\rm lat}^{\rm C}$
 at $z=121.1\,$nm is equal to 24\%. At $z=124.7$, 128.6, and
137.3\,nm the relative errors are 15\%, 28\%, and 16\%, respectively.
The nonmonotonous dependence of the error on $z$ is caused by the
random  error.

The experimental data were compared with computations of the
$F_{\rm lat}^{\rm C}$ within the scattering approach. The geometry of a
perfectly shaped sphere near a flat plate was treated using the PFA
(this introduces an error of less than $z/R\approx 0.01$\% \cite{2})
and the scattering from sinusoidal corrugations was treated precisely
using the Rayleigh theory \cite{12} (see also \cite{7}).
The dielectric properties
of Au were described by the generalized plasma-like permittivity with
the plasma frequency $\omega_p=9.0\,$eV (see \cite{13} for
the parameters of oscillators). The influence of
the surface roughness on the
lateral Casimir force was shown to be negligibly small \cite{9}.
The reason is that the rare tall crystals are distributed
nonperiodically and do not contribute to the lateral force.
The stochastic roughness here has a variance of about 3\,nm 
and contributes less than 1\% of the force.

The lateral Casimir force between a sinusoidally corrugated grating and
a sinusoidally corrugated sphere at $T=300\,$K in thermal equilibrium
was computed using the formula
\begin{eqnarray}
&&
F_{\rm lat}^{C}(z,\varphi)=\frac{4k_BTR}{\Lambda}
\int_{z}^{\infty}\!\!\!dz^{\prime}
\sum_{l=0}^{\infty}{\vphantom{\sum}}^{\prime}
\frac{\partial}{\partial\varphi}
\int_{0}^{\infty}\!\!\!dk_y\int_{0}^{\pi/\Lambda}\!\!\!dk_x
\nonumber \\
&&~~~~~~~
\times\ln\,{\rm det}\left[I-R_1({\mbox{\boldmath$k$}}_{\bot},{i}\xi_l)
R_2({\mbox{\boldmath$k$}}_{\bot},{i}\xi_l,z^{\prime})\right].
\label{eq4}
\end{eqnarray}
\noindent
Here $k_B$ is the Boltzmann constant,
${\mbox{\boldmath$k$}}_{\bot}=(k_x,k_y)$ is the wave vector projection
on the plane of the grating, $\xi_l=2\pi k_BTl/\hbar$, $l=0,\,1,\,2,\,\ldots$
are the Matsubara frequencies,
the prime near the summation sign adds a factor
1/2 to the term with $l=0$, and $R_1$ and $R_2$ are the reflection matrices
from each of
the two parallel gratings at the mean separation $z^{\prime}$
(a similar technique was used in \cite{7} for the Casimir energy at
$T=0$; alternative techniques for the Casimir energy in nonflat
geometries were suggested in \cite{14,15}).
To obtain $R_1$ ($R_2$) one needs the solution of
Maxwell equations in the entire space when  the
other grating is absent.  For the
corrugated area, Maxwell equations can be rewritten as a system of the
first-order ordinary differential equations for the components of the
electromagnetic field parallel to the plane of the gratings. In our case of
sinusoidal gratings the numerical solution of this system was based on explicit
Runge-Kutta formulas of orders 4 and 5, the Dormand-Prince pair, and was
performed using the Matlab package.
Outside the corrugation area the solutions are
given by the Rayleigh expansions.   After matching of the solutions
in different regions one determines the coefficients of the Rayleigh
expansions. Then $R_1$ and $R_2$ are expressed in terms of
these coefficients for the reflection of $E_y$ and $H_y$ components of
the electromagnetic field from the respective gratings.
As an example, in Fig.~2
the computational results for $F_{\rm lat}^{C}$ as a
function of $x/\Lambda$ at $z=124.7\,$nm are shown as a solid line
which is in a very good agreement with the data with no fitting parameters.
 Both the dots and the theoretical line
demonstrate that the lateral Casimir force is asymmetric
(the dependence of the
$F_{\rm lat}^{C}$ on $\varphi$ is strictly sinusoidal only
if the calculation is
restricted to the lowest order $\sim A_1A_2$
\cite{5,9}).
The true asymmetry of the lateral force is obvious even without
the theory curve because the average shift of
each maximum point from the midpoint
of two adjacent minima is $(0.12\pm 0.02)\Lambda$.

The computational results for the maximum value of
$F_{\rm lat}^{C}$ versus $z$ are shown in Fig.~3 as the solid
line which is in an excellent agreement with the experimental data.
For comparison purposes the dashed line in Fig.~3 shows the maximum values
of $F_{\rm lat}^{C}$ computed using the PFA applied not only to the
sphere-plate configuration [as in Eq.~(\ref{eq4})]
but to sinusoidal corrugations
as well. To find the Casimir energy between two
corrugated gratings, we have used the standard Lifshitz formula
which includes the skin depth effects, but
replaced separation $z$ with
$z+A_2\sin(2\pi x^{\prime}/\Lambda+\varphi)-
A_1\sin(2\pi x^{\prime}/\Lambda)$
and made an averaging over $x^{\prime}$.
As can be seen in Fig.~3, the PFA force amplitudes are significantly
larger than
both the experimental data and the exact computational results.
The deviation increases from about 18\% at $z=120\,$nm to
60\% at $z=180\,$nm.
This is explained by the fact that for the shorter $\Lambda$
used in this work the additive approaches are not applicable
\cite{5,6}.

To conclude, we have performed the measurement of the lateral Casimir
force between sinusoidally corrugated surfaces of small period and
compared the obtained data with
the recently developed exact theory taking into
account real material properties of the boundary metal. This
has allowed the first
demonstration of the asymmetry of the lateral Casimir force and has
permitted to quantify deviations from the PFA. The obtained results
enhance the capabilities for the application of the Casimir effect in
nanotechnology, specifically for frictionless transmission of
lateral motion \cite{19n,20n},
and achieve a better ubderstanding of dispersion forces between
nonplanar boundaries.

This work was supported by the NSF Grant No.\ PHY0653657
(measurement of the lateral Casimir force) and DOE Grant
No.\ DE-FG02-04ER46131 (calculation of the electric and Casimir
forces). G.L.K.\ and V.M.M.\ were also supported by the DFG Grant
GE 696/9--1.
V.N.M.\  was supported by the
Grants RNP 2.1.1/1575 and RFBR 07--01--00692--a.


\pagebreak
\begin{figure*}[h]
\caption{(Color online) Schematic of the experimental setup
(see text for further details).
Insertion shows the imprinted corrugations on the second
sphere. The lighter color shows higher points and hence
demonstrates the sphericity of the imprinted surface.}
\end{figure*}
\begin{figure*}[h]
\caption{(Color online) The phase dependence of the lateral Casimir force.
The measurement data are shown as dots. The solid line is
the exact theory.
}
\end{figure*}
\begin{figure*}[h]
\caption{The maximum values of the measured lateral Casimir
force are shown as crosses. The solid and dashed lines are
the predictions of the exact theory and the PFA, respectively.}
\end{figure*}

\end{document}